\documentclass[]{elsart}

\usepackage{graphicx}

\begin{document}

\begin{frontmatter}

\title {Epidemic spreading in lattice-embedded scale-free networks}

\author{Xin-Jian Xu$^{a}$\corauthref{cor}},
\ead{xinjxu@fis.ua.pt}
\corauth[cor]{Corresponding author.}
\author{Zhi-Xi Wu$^{b}$},
\author{Guanrong Chen$^{c}$}

\address{$^{a}$Departamento de F\'{i}sica da Universidade de Aveiro, 3810-193 Aveiro, Portugal}
\address{$^{b}$Institute of Theoretical Physics, Lanzhou University, Lanzhou Gansu 730000, China}
\address{$^{c}$Department of Electronic Engineering, City University of Hong Kong, Kowloon, Hong Kong SAR, China}

\begin{abstract}
We study geographical effects on the spread of diseases in
lattice-embedded scale-free networks. The geographical structure
is represented by the connecting probability of two nodes that is
related to the Euclidean distance between them in the lattice. By
studying the standard Susceptible-Infected model, we found that
the geographical structure has great influences on the temporal
behavior of epidemic outbreaks and the propagation in the
underlying network: the more geographically constrained the
network is, the more smoothly the epidemic spreads, which is different
from the clearly hierarchical dynamics that the
infection pervades the networks in a progressive cascade across
smaller-degree classes in Barab\'{a}si-Albert scale-free networks.
\end{abstract}


\end{frontmatter}


\section{Introduction}

The classical mathematical approach to disease spreading either
ignores the population structure or treats populations as
distributed in a regular medium \cite{Murray}. However, it has
been suggested recently that many social, biological, and
communication systems possess two universal characters, the
small-world effect \cite{Watts} and the scale-free property
\cite{Barabasi}, which can be described by complex networks whose
nodes represent individuals and links represent the interactions
among them \cite{Albert}. In view of the wide occurrence of
complex networks in nature, it is important to study the effects
of topological structures on the dynamics of epidemic spreading.
Pioneering works \cite{Cohen_1,May,Kuperman,Pastor_1} have
given some valuable insights of that: for small-world networks,
there are critical thresholds below which infectious diseases will
eventually die out; on the contrary, even infections with low
spreading rates will prevail over the entire population in
scale-free networks, which radically changes many of the
conclusions drawn in classic epidemic modelling. Furthermore, it
was observed that the heterogeneity of a population network in
which the epidemic spreads may have noticeable effects on the
evolution of the epidemic as well as the corresponding
immunization strategies
\cite{Pastor_1,Cohen_2,Barthelemy_1}.

In many real systems, however, individuals are often embedded in a
Euclidean geographical space and the interactions among them
usually depend on their spatial distances \cite{Yook}. Also, it
has been proved that the characteristic distance plays a crucial
role in the phenomena taking place in the system
\cite{Rozenfeld,Warren,Huang,Crepey}. Thus, it is natural to study
complex networks with geographical properties. Rozenfeld \emph{et
al.} considered that the spatial distance can affect the
connection between nodes and proposed a lattice-embedded
scale-free network (LESFN) model \cite{Rozenfeld} to account for
geographical effects. Based on a natural principle of minimizing
the total length of links in the system, the scale-free networks
can be embedded in a Euclidean space without additional external
exponents. Since distributions of individuals in social networks
are always dependent on their spatial locations, the influence of
geographical structures on epidemic spreading is of high
importance, but up to now it has rarely been studied except for
\cite{Small,Colizza}.

In this paper, we study the standard Susceptible-Infected (SI)
model on the LESFNs, trying to understand how the geographical
structure affects the dynamical process of epidemic spreading.
Especially, we consider the temporal behavior of epidemic
outbreaks and the propagation in the underlying network. It is
found that the geographical structure plays an important role in
the epidemic outbreaks and the propagation of diseases.

\section{The model}

\begin{figure}
\includegraphics[width=\columnwidth]{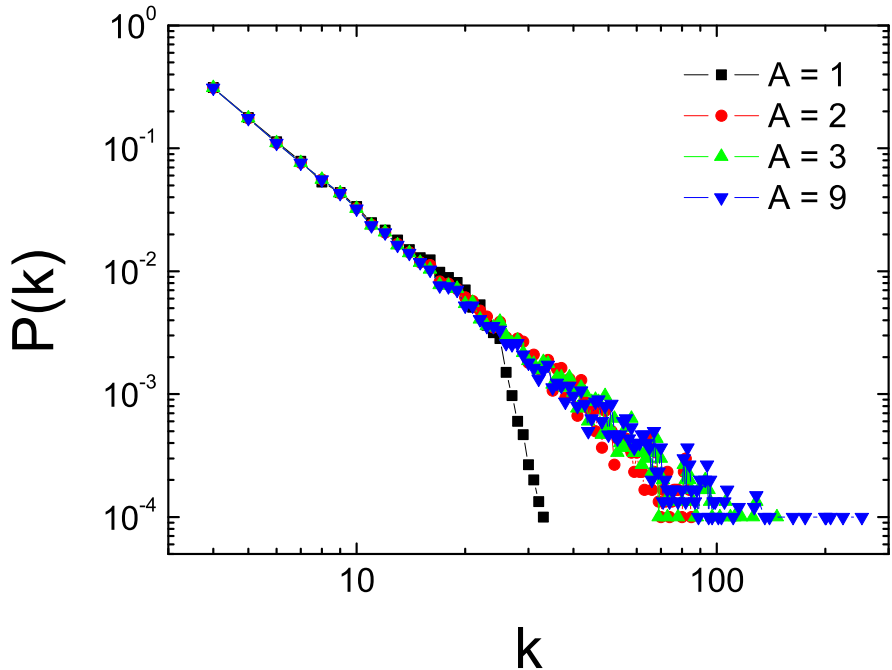}

\caption{(Color online) Degree distribution of an LESFN model with
$N=10000$, $\gamma=2.5$, and for different values of $A$. The
territory parameter $A$ controls the influence of the geographical
distance on the network structure.}\label{pk}
\end{figure}

According to Ref. \cite{Rozenfeld}, the LESFN is generated as
follows: (i) a lattice of size $N=L \times L$ with periodic
boundary conditions is assumed, upon which the network will be
embedded; (ii) for each site, a preset degree $k$ is assigned
taken from the scale-free distribution, $P(k) \sim k^{-\gamma}$,
$m<k<K$; (iii) a node (say $i$, with degree $k_{i}$) is picked out
randomly and connected to its closest neighbors until its degree
quota $k_{i}$ is realized or until all sites up to a distance,
$r(k_{i})=A\sqrt{k_{i}}$, have been explored. Duplicate
connections are avoided. Here, $r(k_{i})$ is the spatial distance
on a Euclidean plane, denoting the characteristic radius of the
region that node $i$ can almost freely connect to the others; (iv)
this process is repeated throughout all the sites on the lattice.
Following this method, networks with $\gamma > 2$ can be
successfully embedded up to a (Euclidean) distance $r(k)$, which
can be made as large as desired depending upon the change of the
territory parameter $A$. Especially, the model turns out to be a
randomly connected scale-free network when $A \rightarrow \infty$
\cite{Newman}. Typical networks with $\gamma=2.5$ resulting from
the embedding method are illustrated in Fig. \ref{pk}. The
power-low degree distributions of the LESFNs achieve their natural
cutoff lengths for $A=2$, $3$, and $9$, respectively. While for
$A=1$, the network ends at some finite-size cutoff length.

In order to study the dynamical evolution of epidemic outbreaks,
we focus on the standard SI model \cite{Murray}. In this model,
individuals have only two discrete states: susceptible (or
healthy) and infected. Each individual is represented by a vertex
of the network and the links are the connections between
individuals along which the infection may spread. There are
initially a number of $I_{0}$ infected nodes and any infected node
can pass the disease to its susceptible neighbors at a spreading
rate $\lambda$. Once a susceptible node is infected, it remains in
this state. The total population (the size of the network) $N$ is
assumed to be constant and, if $S(t)$ and $I(t)$ are the numbers
of susceptible and infected individuals at time $t$, respectively,
then $N=S(t)+I(t)$. In spite of its simplicity, the SI model is a
good approximation for studying the early dynamics of epidemic
outbreaks \cite{Murray,Barthelemy_1} and assessing the effects of
the underlying topologies on the spreading dynamics
\cite{Crepey,Barthelemy_2}.

\section{Simulation results}

We have performed Monte-Carlo (MC) simulations of the SI model
with synchronously updating on the LESFNs. With relevance to
empirical evidence that many networks are characterized by a
power-low distribution with $2<\gamma<3$ \cite{Albert}, we set
$\gamma=2.5$ in the present work. Initially, we select one node
randomly and assume it is infected. The disease will spread
throughout the network and the dynamical process is controlled by
the topology of the underlying network.

\begin{figure}
\includegraphics[width=\columnwidth]{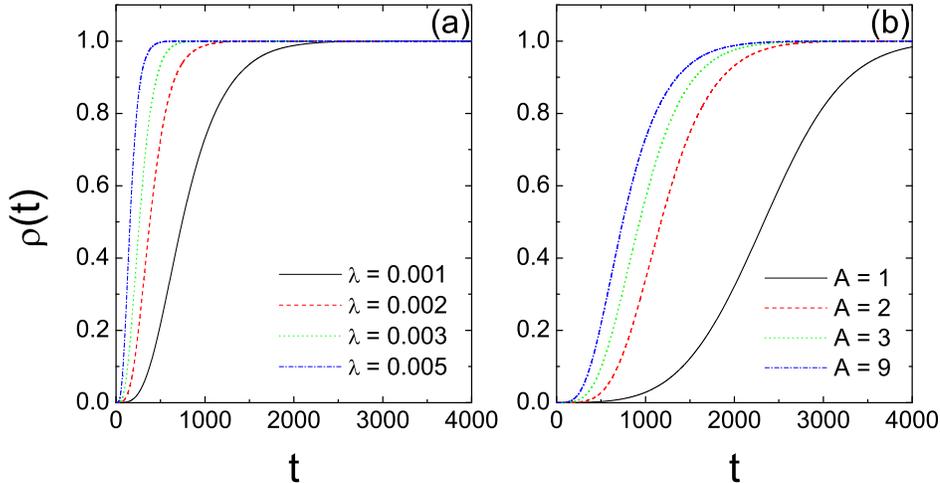}

\caption{(Color online) Density of infected individuals versus MC
time in the LESFNs for different values of $\lambda$ and $A$. (a)
$A=9$, $\lambda=0.001$ (solid line), $0.002$ (dash line), $0.003$
(dot line), and $0.005$ (dash-dot line), respectively. (b)
$\lambda=0.001$, $A=1$ (solid line), $2$ (dash line), $3$ (dot
line), and $4$ (dash-dot line), respectively. The network size is
$N=10000$. All the plots were averaged over $500$
experiments.}\label{it}
\end{figure}

\begin{figure}
\includegraphics[width=\columnwidth]{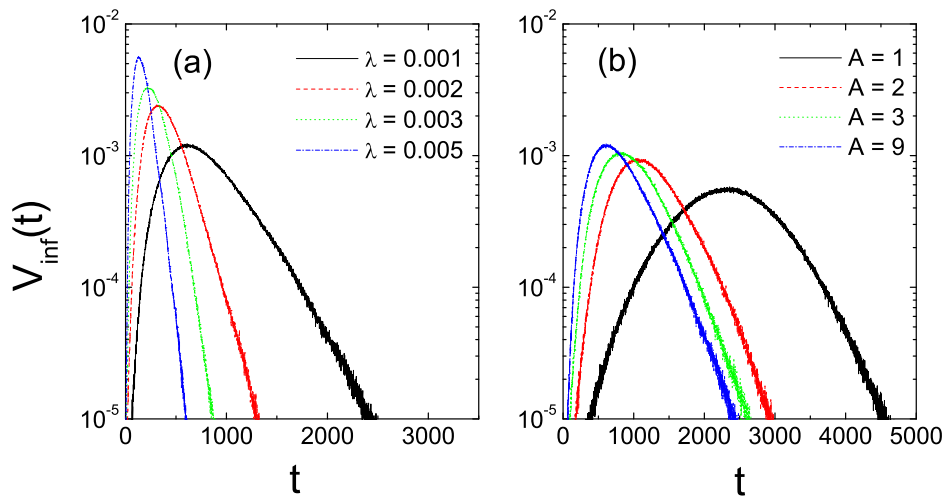}

\caption{(Color online) Linear-log plots of spreading velocity in
the LESFNs for different values of $\lambda$ and $A$. The network
size is $N=10000$. All the curves were averaged over $500$
realizations. The symbols are the same as in Fig.
\ref{it}.}\label{vt}
\end{figure}

In Fig. \ref{it}, we report the temporal behavior of outbreaks in
the LESFNs. The density of infected individuals, $\rho(t)$
($=I(t)/N$), is computed over $500$ realizations of the dynamics.
Consistent with the definition of the model, all the individuals
will be infected in the long-time limit, i.e., $\lim_{t
\rightarrow \infty}\rho(t)=1$, since infected individuals remain
unchanged during the evolution. In Fig. \ref{it}(a), for the given
connecting region ($A=9$), the greater the spreading rate
$\lambda$ is, the more quickly the infection spreads. Fig.
\ref{it}(b) shows the dependence of the infectious prevalence
$\rho(t)$ on the territory parameter $A$, while the spreading rate
is fixed at $\lambda=0.001$. As $A$ increases from $1$ to $9$, the
average spatial length of edges increases \cite{Rozenfeld};
namely, the nodes have larger probabilities to connect to more
other nodes, which therefore leads to a faster spread of the
infection.

To better understand the virus propagation in the population, we
study in detail the spreading velocity, written as \cite{Wu}
\begin{equation}
V_{inf}(t)=\frac{d\rho(t)}{dt} \approx \rho(t)-\rho(t-1).
\label{vinf}
\end{equation}
We count the number of newly infected nodes at each time step and
report the spreading velocity in Fig. \ref{vt}. Apparently, the
spreading velocity goes up to a peak quickly and leaves very short
response time for us to develop control measures. Before the
outbreaks of the infection, the number of infected individuals is
very small, lasting for a very long time during the outbreak, and
the number of susceptible individuals is very small. Thus, when
$t$ is very small (or very large), the spreading velocity is close
to zero. In the case of $A=9$ (Fig. \ref{vt}(a)), all plots show
an exponential decay in the long-time propagation. These results
are weakened for a small circular connecting probability (in
particular, for $A=1$), as shown in Fig. \ref{vt}(b), where the
disease spreads in a relatively low velocity, which therefore
slows down the outbreaks.

\begin{figure}
\includegraphics[width=\columnwidth]{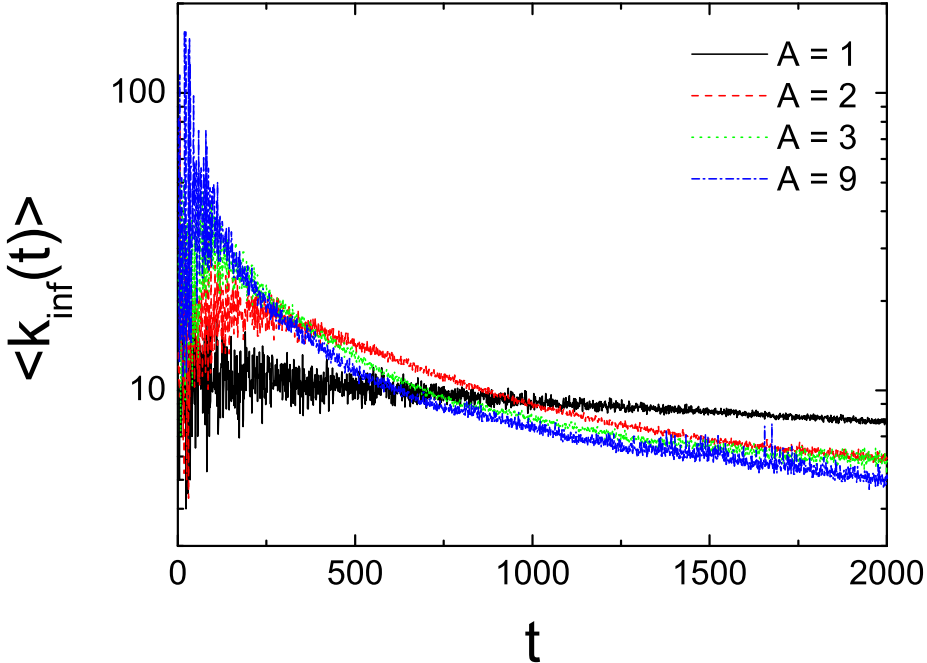}

\caption{(Color online) Temporal behavior of the average degree of
the newly infected nodes for the SI model outbreaks in the LESFNs
with $N=10000$ and $\lambda=0.001$. The data are averaged over
$500$ dynamical processes.}\label{kinft}
\end{figure}

A more precise characterization of the epidemic diffusion through
such a network can be achieved by studying the average degree of
the newly infected individuals at time $t$ \cite{Barthelemy_1},
given by
\begin{equation}
\langle k_{inf}(t) \rangle
=\frac{\sum_{k}kI_{k}(t)-\sum_{k}kI_{k}(t-1)}{I(t)-I(t-1)},
\label{kinf}
\end{equation}
where $I_{k}(t)$ is the number of infected nodes with degree $k$.
In Fig. \ref{kinft}, we plot the temporal behavior of $k_{inf}(t)$
for the SI model with $\lambda=0.001$ in the LESFNs. Different
from the clearly hierarchical dynamics in Barab\'{a}si-Albert
scale-free networks \cite{Barthelemy_1}, in which the infection
pervades the networks in a progressive cascade across
smaller-degree classes, epidemic spreads slowly from higher- to
lower-degree classes in the LESFNs. The smaller the territory
parameter $A$ is, the more smoothly the infection propagates.
Especially, strong global oscillations arise in the initial time
region, which implies that the geographical structure plays an
important role in early epidemic spreading, independent of the
degrees of the infected nodes.

\begin{figure}
\includegraphics[width=\columnwidth]{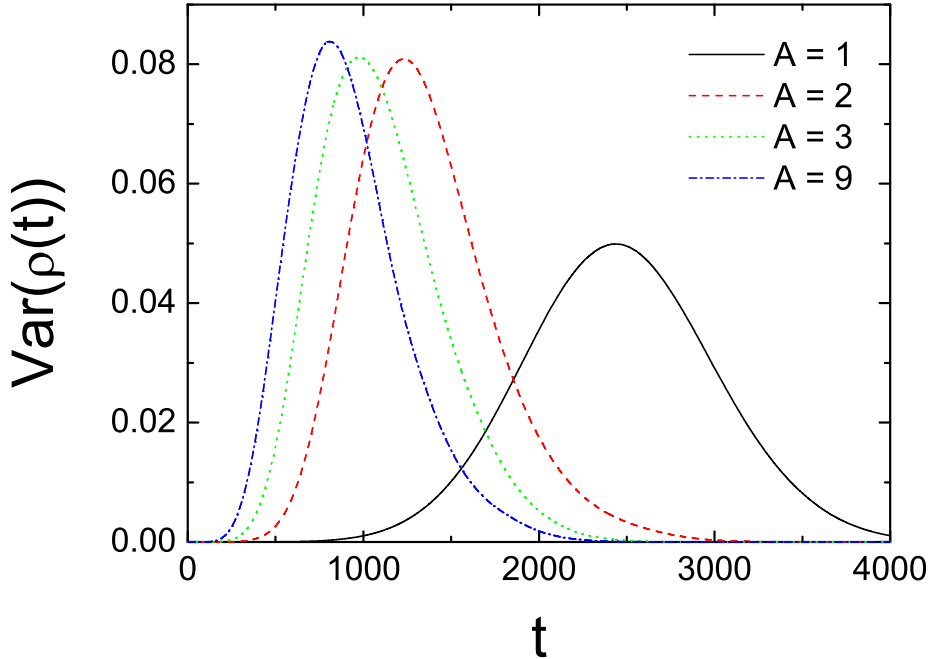}

\caption{(Color online) Temporal behavior of the variance of the
density of infected nodes in the LESFNs for different values of
$A$. The results are obtained for $\lambda=0.001$ and on networks
of size $N=10000$.}\label{vart}
\end{figure}

Since the intrinsic stochasticity of the epidemic spreading makes
each realization unique \cite{Crepey,Colizza}, it is valuable to
analyze the statistical fluctuations around the average behavior
for assessing simulation results with respect to real outbreaks.
We measure the variance of the prevalence, defined by
\begin{equation}
Var(\rho(t))=E[(\rho(t)-\overline{\rho}(t))^{2}].\label{var}
\end{equation}
In order to evaluate this quantity, we have performed $500$
independent runs with different configurations of the intrinsic
frequencies as well as different network realizations. Fig. \ref
{vart} displays the time evolution of $Var(\rho(t))$ for LESFNs
with different values of $A$. Since the initial prevalence is
fixed and is the same for all instances, $Var(\rho(t))$ is
initially equal to zero and can only increase. At very large time,
almost all nodes are infected, implying that
$\lim_{t\rightarrow\infty}Var(\rho(t))=0$. Compared with Fig.
\ref{vt}(b), one can easily find that time regions in which the
fluctuations are maximal are the same as that in which the
spreading velocities are the fastest. Moreover, an important
feature is the rough symmetry of curves, regardless of $A=9$ (more
randomly scale-free) or $A=1$ (more locally constrained) and the
time regimes in which the fluctuations are maximal corresponding
to a small diversity of the degrees of the infected nodes. This is
also different from that in Barab\'{a}si-Albert scale-free
networks \cite {Crepey}.

\section{Conclusions}

We have studied geographical effects on the spreading phenomena in
lattice-embedded scale-free networks, in which a territory
parameter $A$ controls the influence of the geographical structure
on the network. We have investigated the temporal behavior of
epidemic outbreaks and found that the spreading velocity reaches a
peak very quickly in the initial infection period and then decays
in an approximately exponential form in a more randomly scale-free
region, which is consistent with previous studies. While the
networks are more graphically constrained, this feature is
apparently weakened. Furthermore, we have studied the propagation
of the infection through different degree classes in the networks.
Different from the clearly hierarchical dynamics in which the
infection pervades the networks in a progressive cascade across
smaller-degree classes in Barab\'{a}si-Albert scale-free networks,
epidemic smoothly spreads from higher- to lower-degree classes in
the LESFNs. Finally, we have analyzed the prevalence fluctuations
around the average epidemic process. A rough symmetry of curves is
found and the time regions in which the fluctuations are maximal
correspond to a small diversity of the degrees of the infected
nodes, which is also different from that observed from
Barab\'{a}si-Albert scale-free networks.

\section{Acknowledgements}

X.-J. Xu acknowledges financial support from FCT (Portugal), Grant
No. SFRH/BPD/30425/2006. G. Chen acknowledges the Hong Kong
Research Grants Council for the CERG Grant CityU 1114/05E.


\begin{thebibliography}{99}

\bibitem{Murray}
J.D. Murray, Mathematical Biology, Springer Verlag, Berlin, 1993;
R.M. Anderson, R.M. May, Infectious Diseases in Humans, Oxford
University Press, Oxford, 1992.

\bibitem{Watts}
D.J. Watts, S.H. Strogatz, Nature 393 (1998) 440.

\bibitem{Barabasi}
A.-L. Barab\'{a}si, R. Albert, Science 286 (1999) 509; A.-L.
Barab\'{a}si, R. Albert, H. Jeong, Physica A 272 (1999) 173.

\bibitem{Albert}
R. Albert, A.-L. Barab\'{a}si, Rev. Mod. Phys. 74 (2002) 47;
S.N. Dorogovtsev, J.F.F. Mendes, Adv. Phys. 51 (2002) 1079;
M.E.J. Newman, SIAM Rev. 45 (2003) 167.

\bibitem{Cohen_1}
R. Cohen, K. Erez, D. ben-Avraham, S. Havlin, Phys. Rev. Lett.
85 (2000) 4626.

\bibitem{May}
R.M. May, A.L. Lloyd, Phys. Rev. E 64 (2001) 066112.

\bibitem{Kuperman}
M. Kuperman, G. Abramson, Phys. Rev. Lett. 86 (2001) 2909.

\bibitem{Pastor_1}
R. Pastor-Satorras, A. Vespignani, Phys. Rev. Lett. 86 (2001)
3200; Phys. Rev. E 63 (2001) 066117.

\bibitem{Cohen_2}
R. Cohen, S. Havlin, D. ben-Avraham, Phys. Rev. Lett. 91
(2003) 247901.

\bibitem{Barthelemy_1}
M. Barth\'{e}lemy, A. Barrat, R. Pastor-Satorras, A.
Vespignani, Phys. Rev. Lett. 92 (2004) 178101.

\bibitem{Yook}
S.-H. Yook, H. Jeong, A.-L. Barab\'{a}si, Proc. Natl. Acad.
Sci. USA 99 (2002) 13382; G. Nemeth, G. Vattay, Phys. Rev. E 67
(2003) 036110; M.T. Gastner, M.E.J. Newman, Eur. Phys. J. B 49
(2006) 247.

\bibitem{Rozenfeld}
A.F. Rozenfeld, R. Cohen, D. ben-Avraham, S. Havlin, Phys.
Rev. Lett. 89 (2002) 218701; D. ben-Avraham, A.F. Rozenfeld, R.
Cohen, S. Havlin, Physica A 330 (2003) 107.

\bibitem{Warren}
C.P. Warren, L.M. Sander, I.M. Sokolov, Phys. Rev. E 66
(2002) 056105.

\bibitem{Huang}
L. Huang, L. Yang, K. Yang, Phys. Rev. E. 73 (2006) 036102.

\bibitem{Crepey}
P. Cr\'{e}pey, F. Alvarez, M. Barth\'{e}lemy, Phys. Rev. E 73
(2006) 046131.

\bibitem{Small}
M. Small, C.K. Tse, Physica A 351 (2005) 499; M. Small, C.K.
Tse, D.M. Walker, Physica D 215 (2006) 146.

\bibitem{Colizza}
V. Colizza, A. Barrat, M. Barth\'{e}lemy, A. Vespignani, Proc.
Natl. Acad. Sci. USA 103 (2006) 2015.

\bibitem{Newman}
M.E.J. Newman, S.H. Strogatz, D.J. Watts, Phys. Rev. E 64 (2001)
026118.

\bibitem{Barthelemy_2}
M. Barth\'{e}lemy, A. Barrat, R. Pastor-Satorras, A.
Vespignani, J. Theor. Biol. 235 (2005) 275.

\bibitem{Wu}
Z.-X. Wu, X.-J. Xu, Y.-H. Wang, Eur. Phys. J. B 45 (2005) 385.

\end{thebibliography}
\end{document}